\documentclass[dvips]{aa}

\usepackage{amsmath} \usepackage{amssymb}\usepackage{graphicx}
\usepackage{txfonts} \usepackage{natbib} \usepackage{multirow}

\bibpunct{(}{)}{;}{a}{}{,}
 
\begin{document}
 
\title{The characteristics of the IR emission features in the spectra
  of Herbig Ae stars: Evidence for chemical evolution}

\titlerunning{Chemical evolution in Herbig Ae stars}

\author{C. Boersma\inst{1} \and J. Bouwman\inst{2} \and
  F. Lahuis\inst{3,4} \and C. van Kerckhoven\inst{5} \and
  A.G.G.M. Tielens\inst{1,6} \and L.B.F.M. Waters\inst{7} \and
  Th. Henning\inst{2}}

\institute{Kapteyn Astronomical Institute, University of Groningen,
  P.O. Box 800, 9700 AV, Groningen, The Netherlands \and Max-Planck
  Institut f\"ur Astronomie, K\"oningstuhl 17, 69117 Heidelberg,
  Germany \and Space Research Organisation Netherlands, P.O.Box 800,
  9700 AV Groningen, The Netherlands \and Leiden Observatory, P.O. Box
  9513, 2300 RA Leiden, The Netherlands \and Instituut voor
  Sterrenkunde, K. U. Leuven, Celestijnlaan 200B, 3001 Heverlee,
  Belgium. \and NASA Ames Research Center, MS 245-3, Moffett Field, CA
  94035, USA. \and Astronomical Institute ``Anton Pannekoek'',
  University of Amsterdam, Kruislaan 403, 1098 SJ Amsterdam, The
  Netherlands}

\offprints{C.Boersma AT astro.rug.nl}

\date{Received January 16, 2006}


\abstract {Infrared (IR) spectra provide a prime tool to study the
  characteristics of polycyclic aromatic hydrocarbon (PAH) molecules
  in regions of star formation. Herbig Ae/Be stars are a class of
  young pre-main sequence stellar objects of intermediate mass. They
  are known to have varying amounts of natal cloud material still
  present in their direct vicinity.} {We characterise the IR emission
  bands, due to fluorescence by PAH molecules, in the spectra of
  Herbig Ae/Be stars and link observed variations to spatial aspects
  of the mid-IR emission.} {We analysed two PAH dominated spectra from
  a sample of 15 Herbig Ae/Be stars observed with the Spitzer Space
  Telescope.} {We derived profiles of the major PAH bands by
  subtracting appropriate continua. The shape and the measured band
  characteristics show pronounced variations between the two Spitzer
  spectra investigated. Those variations parallel those found between
  three infrared space observatory (ISO) spectra of other,
  well-studied, Herbig Ae/Be stars. The derived profiles are compared
  to those from a broad sample of sources, including reflection
  nebulae, planetary nebulae, \ion{H}{ii} regions, young stellar
  objects, evolved stars and galaxies. The Spitzer and ISO spectra
  exhibit characteristics commonly interpreted respectively as
  interstellar matter-like (ISM), non-ISM-like, or a combination of
  the two.} {We argue that the PAH emission detected from the sources
  exhibiting a combination of ISM-like and non-ISM-like
  characteristics indicates the presence of two dissimilar, spatially
  separated, PAH families. As the shape of the individual PAH band
  profiles reflects the composition of the PAH molecules involved,
  this demonstrates that PAHs in subsequent, evolutionary linked
  stages of star formation are different from those in the general
  ISM, implying active chemistry. None of the detected PAH emission
  can be associated with the (unresolved) disk and is thus associated
  with the circumstellar (natal) cloud. This implies that chemical
  changes may already occur in the (collapsing?)  natal cloud and not
  necessarily in the disk.}


\maketitle

\section{Introduction}
\label{sec:sec1}

Herbig Ae/Be stars are a class of young stellar objects
\citep{1960ApJS....4..337H} originally limited to pre-main sequence
stars of intermediate mass (2 - 8 M$_{\odot}$) with spectral class A
or B with associated nebulosity and showing emission lines. Over the
years, however, the definition has been somewhat relaxed, e.g. the
necessity for associated nebulosity \citep{1998A&A...331..211M}.

The formation and evolution of isolated, low-mass stars, is relatively
well understood \citep{1987ARA&A..25...23S}. However, the genesis of
intermediate and high-mass stars remains more elusive. For
intermediate mass, Herbig Ae/Be stars, it is thought that the
formation scheme is more or less a scaled up version of that for low-
mass stars. The picture that has thus emerged is that stars form in
large molecular clouds and through successive stages of fragmentation,
gravitational collapse, and disk accretion, accompanied by bipolar
outflows, the star reaches the main sequence. During the later stages
of the \emph{pre}-main sequence evolution (0.5 - 10 Myr), the
increased stellar winds from the central object will have removed most
of the surrounding molecular cloud material, leaving only the remnant,
passive, accretion disk ($R_{\text{disk}}\sim 100 \text{AU},\
M_{\text{disk}}\sim5.6\cdot10^{-4} - 1.5\cdot10^{-1}\
\mathrm{M}_{\odot}$; \citealt{2004A&A...427..179H},
\citealt{2004A&A...422..621A}, \citealt{1997ISI...119.W}). This disk -
believed to be the site of planet formation - shows up in polarised
light and as a strong excess in infrared spectra.

With the 1995 launch of the European Space Agency's Infrared Space
Observatory (ESA ISO, \citealt{1996A&A...315L..27K}), it became
possible for the first time to study the infrared spectrum of Herbig
Ae/Be stars in detail. Those studies revealed a large variety in dust
composition, both carbonaceous and silicate in nature, and in dust
properties \citep{2000A&A...357..325V, 2000A&A...360..213B,
  1996A&A...315L.245W}. Specifically, strong infrared (IR) features
with satellite features and emission plateaus around 3.3, 6.2, 7.6,
7.8, 8.6, 11.2, and 12.7 micron, associated with polycyclic aromatic
hydrocarbons (PAHs), were apparent in many of these objects. PAHs,
large molecules of many fused aromatic rings, fluoresce in the
infrared upon the absorption of a single visible or UV photon
\citep{1984A&A...137L...5L, 1985ApJ...299L..93C, 1989ApJS...71..733A}.
Driven by the dispersal of the disk and the formation of a planetary
system, protoplanetary disks are continuously evolving. Geometry, gas,
and dust content; gas and dust processing; and transport and mixing
are important issues concerning these disks. The diversity of spectral
features and their variations within and from source-to-source reflect
these issues. However, the interrelationship of the variance in dust
composition, the characteristics of the disk, and the global system is
not well understood.

Over the years, PAHs have proven to be an ideal probe for many of
these properties, e.g., disk geometry \citep{2001A&A...365..476M}. The
presence of (large) PAH molecules influences many aspects in
protoplanetary systems. These aspects include: (surface) chemistry,
due to the large surface area of PAHs; heating and cooling through
photo-electric ejection; infrared emission; gas-grain collisions; and
the charge balance, which in its turn influences the equilibrium state
of chemical reactions.

Today, the superior sensitivity of NASA's Spitzer Space Telescope
\citep{2004ApJS..154....1W} allows us to investigate protoplanetary
disk systems in better detail than before. Spectra of 15 nearby Herbig
Ae/Be stars (Sect. \ref{sec:sec2}), obtained with the Infrared
Spectrograph\footnote{The IRS was a collaborative venture between
  Cornell University and Ball Aerospace Corporation funded by NASA
  through the Jet Propulsion Laboratory and Ames Research Center.}
\citep[IRS,][]{2004ApJS..154...18H}, are investigated for emission due
to PAHs (Sect. \ref{sec:sec3}). We characterise the individual
profiles for the two PAH-dominated spectra found in the sample (Sects.
\ref{subsec:sec3subsec1} \& \ref{subsec:sec3subsec2}) and, by using
parallels between the morphology of the stellar systems of these
sources and those of three well-studied systems, explore possible
causes for the observed variations in band profiles (Sec.
\ref{sec:sec4}). Similarities between the 6.2 and ``7.7'' $\mu$m band
profiles are treated (Sect. \ref{subsec:sec4subsec1}) and the extended
nature of the PAH emission is investigated (Sect.
\ref{subsec:sec4subsec2}). The origin of the variations in peak
position is discussed (Sect. \ref{subsec:sec4subsec3}) and from this,
implications about the onset of the chemical modification of PAHs -
disk and/or envelope - are inferred (Sect. \ref{subsec:sec4subsec4}).
We finish by summarising our findings (Sect. \ref{sec:sec5}).

\section{Data}
\label{sec:sec2}

The IRS instrument on board Spitzer was used to obtain the spectra of
36 Herbig Ae/Be \citep{1960ApJS....4..337H} stars. We selected the
stars from a list of Herbig Ae/Be stars compiled by
\citet{1998A&A...331..211M} and they complement the 26 Herbig Ae/Be
stars observed in the Spitzer guaranteed time observations and legacy
programs.

For 15 genuine Herbig Ae/Be stars, we have access to the appropriate
wavelength region for detecting emission due to PAHs. Those stars were
observed using the short-high and short-low (SH, SL resp.) modules in
high accuracy peak-up mode ($\sigma\sim0.14^{\prime\prime}$),
providing spectra at $R\sim 600$ for the high-resolution module and
$R\sim 64 - 128$ for the low-resolution module, covering wavelengths
from 9.9 - 19.6 and 5.2 - 14.5 $\mu$m for the SH and SL modules,
respectively.

The Spitzer Science Center (SSC) processed the data with the science
data pipeline version S13.2.0. In order to make more accurate
corrections for background emission than possible with the standard
pipeline,

we adopted the extraction techniques developed by the formation and
evolution of protoplanetary systems (FEPS) and cores to disks (C2D)
Spitzer legacy teams (Bouwman in prep., \citealt{2007PhDT.........2L}
and c2d-IRS delivery
documentation\footnote{http://data.spitzer.caltech.edu/popular/c2d/\\
  20061201\_enhanced\_v1/Documents/C2D\_Spectr\_Expl\_Supp.pdf}).
Extraction of the one-dimensional spectra are done on the intermediate
\texttt{droopres} and \texttt{rsc} products that are provided by the
SSC. These data have most of the \emph{Spitzer-specific} artifacts
removed. For the \texttt{rsc} products this includes straylight and
crosstalk corrections, but not flat-fielding. The absolute flux
calibration is done using dedicated spectral response functions,
derived from IRS spectra of a suite of calibrators using Cohen and
MARCS stellar models provided by the SSC.

The high-resolution spectra are extracted from the \texttt{rsc}
products using an optimal source profile extraction developed by the
c2d team. The optimal extraction uses an analytical point spread
function (PSF) defined from high $S/N$ sky corrected calibrator data.
A simultaneous cross-disperion source profile and extended emission
fit is made to the combined dither data. For the extended component,
the cross-dispersion profiles of the flat-field images are used. The
applied source profile is optimised by comparing the normalised (in
cross-dispersion direction) and collapsed (in dispersion/wavelength
direction) science source profile with the average profile of the
calibrator (point) sources. Bad-pixels are identified and excluded
from the profile fits. The optimal extraction allows us to correct the
spectrum for local extended emission (see, e.g.,
\citealt{2006A&A...459..545G,2007ApJ...665..492L}), which is crucial
in confused regions where the use of separate sky observations is
limited.

The low-resolution spectra are extracted from the \texttt{droopres}
products using a 6.0 pixel fixed-width aperture. Background
corrections are made by subtracting the two-dimensional spectral
images from the associated nod positions. Besides stray light, this
also corrects for anomalous dark current effects. Identified bad
pixels are interpolated from the surrounding perimeter. The optimal
source extraction is used to derive a possible residual extended
emission component in the extraction aperture of the low-resolution
order 1 spectra (see Fig. \ref{fig:fig8}. Due to the severe
undersampling of the spectral images in the spatial direction, no
source profile fits are made for order 2 and 3 and hence no estimate
of the extended emission is derived. Flux offsets between orders
(particular in the SL module) and modules are apparent. They are
probably related to pointing errors and are likely to have introduced
low level fringing and small slope changes. For the SL module, we
developed a procedure to estimate the pointing errors and correct for
them. This procedure minimises the total variance between the flux
levels of the individual cycles, nod positions and orders, applying
wavelength dependent correction factors to the flux for each relative
offset. The procedure simultaneously corrects for fringes. For the
high-resolution modules, fringes are removed using the {\sc
  irsfringe} package \citep{2003cdsf.conf..335L}. The
SMART\footnote{SMART was developed by the IRS Team at Cornell
  University and is available through the Spitzer Science Center at
  Caltech.  \citep{2004PASP..116..975H} software package has been
  utilised to further reduce and analyse the data.} A mismatch between
flux levels of the SL and SH modules is corrected by multiplying the
SH flux levels with a constant, which is determined in the overlap
region and no adjustment for any differences in tilt for the two
sub-spectra is made. As the final reduction steps, outliers are
removed and, in regions of spectral overlap, the orders are spliced to
form a continuous profile. For rising spectra this splicing may
introduce small artifacts in the overlap regions, most notably near
7.5 $\mu$m. This is probably related to weak order leakage effects
introduced by the spectral response function, which has been
determined using only declining stellar spectra. In the case of
extended sources, artifacts may arise due to small differences in
aperture size between orders and modules. The relative errors are
dominated by the intrinsic (Gaussian) noise in each spectral data
point, not by uncertainties in the calibration
\citep{2005FEPS.book.....H}. The intrinsic noise is characterised by
the deviations found between subsequent nod positions and cycles and
are about 2 percent across an order. Between orders it is estimated to
be about 5 percent and between modules about 8 percent.

\section{Analysis}
\label{sec:sec3}

In order to investigate the emission due to PAHs we roughly classify
the spectra of the 15 sources by their dominant spectral
component. Table \ref{tab:tab1} presents our classifications.

\begin{table}[htbp]
  \centering
  \caption{Overview of the 15 genuine Herbig Ae/Be stars with their dominant spectral component (DC).}
  \label{tab:tab1}
  \begin{minipage}{\linewidth}
    \centering
    \begin{tabular}{lc|lc}
      \hline
      \hline
      Source          & DC$^\dagger$ & Object   & DC$^\dagger$ \\
      \hline
      HD58647         & Si      & HD38120  & SP \\ 
      HD35929         & Si      & HD72106  & SP \\ 
      HD37258         & Si      & HD85567  & SP \\ 
      {\bf HD36917}  & {\bf P} & HD95881  & SP \\ 
      {\bf HD37411}  & {\bf P} & HD142527 & SP \\  
      HD36112         & SP      & HD190073 & SP \\      
      HD37357         & SP      & HD244604 & SP \\ 
      HD37806         & SP      &               \\          
      \hline
    \end{tabular}
    \begin{flushleft}
      $^\dagger$ Si : Silicate dominated, P : PAH dominated, SP : Silicates and PAHs
    \end{flushleft}
  \end{minipage}
\end{table}

To avoid the problem of separating the PAH bands from other dust
components, we focus here on the analysis of the two PAH dominated
spectra in the sample. In forthcoming papers the spectra dominated by
silicate- or with mixed silicate-PAH emission features will be
analysed (Bouwman et al. in prep.; Boersma et al. in prep.)

\begin{figure}[htb]
  \centering
  \includegraphics[angle=0, width=\linewidth]{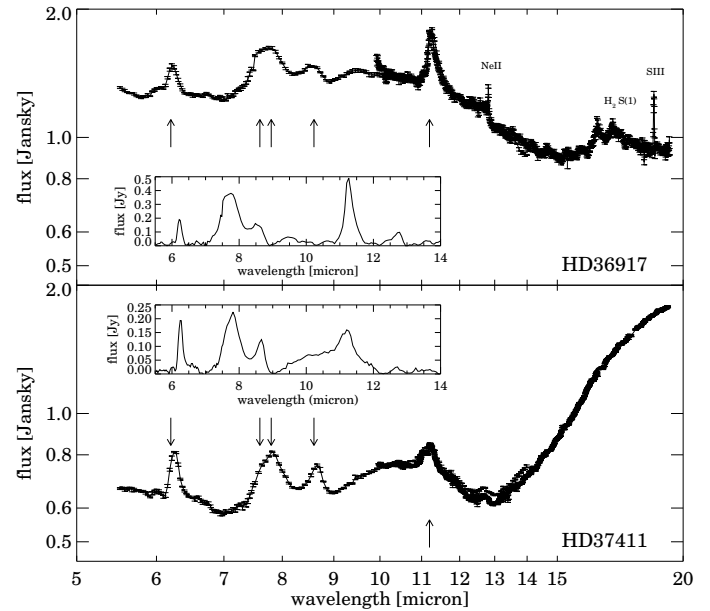}
  \caption{Fully-reduced SL and SH spectra for HD36917 and HD37411. The arrows indicate the positions of the PAH bands of interest. The inset displays the residual (SL) spectrum when approximating the continuum with a single spline.}
  \label{fig:fig1}
\end{figure}

The Herbig Ae/Be stars associated with these two spectra are
\object{HD36917} and \object{HD37411}. Both stars are located south of
the Orion belt in the OB1c association, of which HD36917 is a
spectroscopic binary \citep{1976PASP...88..712L}. The shape of its
SED, with its drop between 2 - 20 $\mu$m and rise beyond 20 $\mu$m,
suggest a cleared (inner) region in its circum-stellar disk (planet?).
The star HD37411 is a visual binary, showing no emission lines. The
optical spectrum of the binary indicates a late type (K-M) companion
(Van den Ancker, unpublished). Table \ref{tab:tab2} summarises the
astrometric data and Table \ref{tab:tab3} the available IRS data.
Figure \ref{fig:fig1} presents the fully reduced Spitzer SL and SH
spectra.

\begin{table*}[hbt]
  \centering
  \caption{Astrometric data for HD36917 and HD37411.}
  \label{tab:tab2}
  \begin{minipage}{\linewidth}
    \centering
    \begin{tabular}{lllllllll}
      \hline
      \hline
      Object   &  $\alpha$ [2000] & $\delta$ [2000] & Region     & $d$[pc] & Ref. & Sp. Type    & Ref. & $A_{\mathrm{V}}$[mag] \\
      \hline
      HD36917  & 05h34m46.98s     & -05d34m14.6s    & Orion OB1c & 510     & 1    & B9.5V + A0V & 2    & 0.49                 \\ 
      HD37411  & 05h38m14.51s     & -05d25m13.3s    & Orion OB1c & 510     & 1    & A0 + K-M    & 3    & 0.00                 \\ 
      \hline                                                                               
    \end{tabular}
    \begin{flushleft}
        1: \citet{1999AJ....117..354D}; 2: \citet{1976PASP...88..712L}; 3: \citet{1998AJ....116.2530G};\\
    \end{flushleft}
  \end{minipage}
\end{table*}

\begin{table}[htb]
  \caption{Available IRS data on HD36917 and HD37411.}
  \label{tab:tab3}
  \begin{minipage}{\linewidth}
  \centering
    \begin{tabular}{lcccc|c}
      \hline
      \hline
               & \multicolumn{3}{c}{Module}              &                    &    \\
      Object   & SL\footnotemark[1] & SH\footnotemark[1] & LH\footnotemark[1] & PU-blue\footnotemark[2] & \#\footnotemark[3] \\
      \hline                                                
      HD36917  & $\surd$ (2)        & $\surd$ (2)        & $\surd$ (3)        &                         & 18                 \\
      HD37411  & $\surd$ (2)        & $\surd$ (3)        & $\surd$ (4)        & $\surd$                 & 22                 \\
      \hline                                                                               
    \end{tabular}
    \begin{flushleft}
      1 : Number of cycles in parentheses, 2 : Peak-up blue image (13.5 - 18.7 $\mu$m) 3 : Total number of spectra
    \end{flushleft}
  \end{minipage}
\end{table}

In order to derive the individual PAH profiles, a continuum is
established. To facilitate comparison with previous studies we adopt
the procedures as outlined by \citet{2002A&A...390.1089P},
\citet{2004ApJ...611..928V}, and \citet{2001A&A...370.1030H}, assuming
PAHs are not significantly contributing to the continuum. In
accordance with these procedures, the continuum for the 6.2 $\mu$m
region is approximated by a single spline; for the 7 - 9 $\mu$m region
a general continuum and an additional plateau component are defined,
where the general continuum is approximated by a single spline, with
points selected between 5 - 6 $\mu$m, 9 - 10 $\mu$m and near 7 $\mu$m.
With additional points near 8.3 $\mu$m the plateau component is fixed.
The continuum in the 11.2 $\mu$m region is approximated by a single
spline. Points between 9 - 10.5, 14.5 - 15.5 and near 11.8 and 13.1
$\mu$m are selected to define a single spline to approximate the
continuum in the 12 $\mu$m region.

\subsection{Results}
\label{subsec:sec3subsec1}

Table \ref{tab:tab4} summarises the identified PAH bands and plateaus
in the continuum subtracted spectra. The spectra of the two sources
show all of the well-known IR emission features at 6.2, 7.7, 8.6, and
11.2 $\mu$m and, in addition, in some sources, the weak bands at 5.25,
5.7, 6.0, 12.0, 12.7, and 16.4 $\mu$m and the broad plateaus
underneath the 6.2, 7.7, and 15 $\mu$m regions. Table \ref{tab:tab5}
presents the spectral properties of the major PAH bands. These
properties were determined fitting a single Gaussian to the band.

Apart from PAH band emission also other, dust and non-dust components
are identified. For instance, the aliphatic hydrocarbon features at
6.85 and 7.25 $\mu$m \citep[][ and refs therein]{2005ApJ...632..956S}
in the SL spectra and the \ion{Ne}{ii}, H$_{2}$ S(1) and \ion{S}{III}
lines at, respectively, 12.8, 17.1, and 18.7 $\mu$m in the SH spectra
of HD36917. Between 9.0 - 13 $\mu$m, a broad emission component,
reminiscent of warm silicate dust, is visible, particularly in
HD37411. In our analyses of the emission feature around 11 $\mu$m we
subtracted this component using a spline fit (see Fig. 4).

\begin{table*}[hbtp]
  \centering
  \caption{Detected PAH band components in HD36917 and HD37411.}
  \label{tab:tab4}
  \begin{tabular}{lccccccccccccc}
    \hline
    \hline
    Source      & 5.25     & 5.7         & 6.0	        & 6.2         & 7.6/7.8   & 8.6	       & 11.0       & 11.2	      & 12.0       & 12.7      & 13.5	     & 16.4       & Plateau \\
    \hline
    HD36917	& $\surd$  & $\surd$ & $\surd$ & $\surd$ & $\surd$ & $\surd$ &	       & $\surd$ & $\surd$ & $\surd$ & $\surd$ & $\surd$ & $\surd$  \\
    HD37411	& $\surd$  &	         & $\surd$ & $\surd$ & $\surd$ & $\surd$ & ?	       & $\surd$ & ?	      & $\surd$ &	               &                &                 \\
    \hline                                                                               
  \end{tabular}
\end{table*}

\begin{table}[hbt]
  \centering
  \caption{Spectral properties of the major PAH bands.}
    \label{tab:tab5}
    \begin{minipage}{\linewidth}
      \centering
      \begin{tabular}{l|l|ll}
        \hline
        \hline
                                              &                           & HD36917 & HD37411\\
        \hline
        \multirow{4}{*}{6.2}                  & position\footnotemark[1]  & 6.22    & 6.25   \\
                                              & FWHM\footnotemark[1]       & 0.27    & 0.15   \\
                                              & EW\footnotemark[2]           & 0.028   & 0.049  \\
                                              & F\footnotemark[3]           & 2.7     & 2.37   \\
        \hline
        \multirow{4}{*}{7.7}                  & position        & 7.63    & 7.78   \\
                                              & FWHM         & 0.42    & 0.43   \\
                                              & EW          & 0.092   & 0.12   \\
                                              & F            & 6.3     & 3.63   \\
        \hline
        \multirow{4}{*}{8.6}                  & position         & 8.58    & 8.64   \\
                                              & FWHM         & 0.33    & 0.25   \\
                                              & EW           & 0.022   & 0.036  \\
                                              & F            & 1.26    & 0.97   \\
        \hline
        \multirow{4}{*}{11.2}                 & position        & 11.29   & 11.21  \\
                                              & FWHM        & 0.31    & 0.38   \\
                                              & EW            & 0.12    & 0.053  \\
                                              & F            & 3.74    & 0.94   \\
        \hline
        \multirow{4}{*}{11.2\footnotemark[4]} & position       & 11.32   & 11.22 \\
                                              & FWHM        & 0.42    & 0.29  \\
                                              & EW          & 0.11    & 0.063 \\
                                              & F            & 4.4     & 1.1   \\
        \hline
        \multirow{4}{*}{12.7}                 & peak         & 12.72   & 12.71  \\
                                              & FWHM          & 0.34    & 0.33   \\
                                              & EW            & 0.026   & 0.013  \\
                                              & F            & 0.67    & 0.16   \\
        \hline
      \end{tabular}
      \begin{flushleft}
        1 : [$\mu$m], 2 : Equivalent width : $\mathbf\int[(band - continuum)/continuum]\mathrm{d}\lambda$\\
        3 : [$10^{-15}\cdot\mathrm{W}\cdot\mathrm{m}^{-2}]$, 4 : Hi-res profile.
      \end{flushleft}
    \end{minipage}
\end{table}

\subsection{Profiles}
\label{subsec:sec3subsec2}

\begin{figure}[htb]
  \centering
  \includegraphics[angle=0, width=\linewidth]{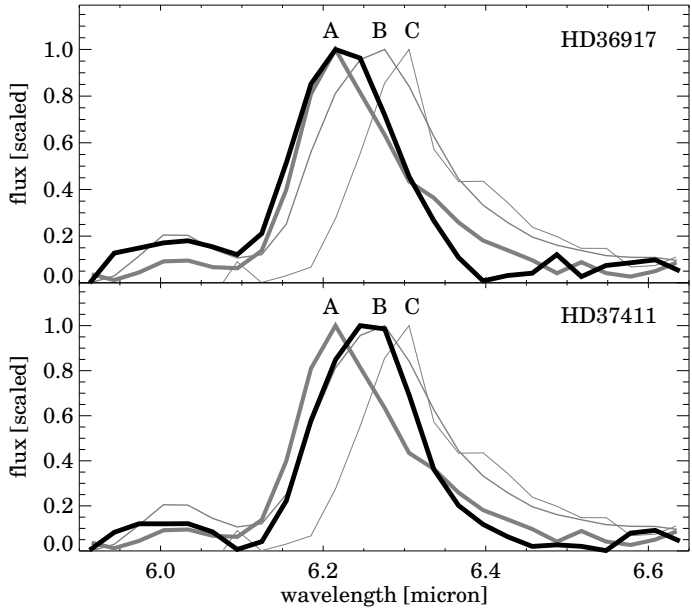}
  \caption{The derived 6.2 $\mu$m profiles (black) after subtraction of the underlying continuum. For comparison, the three characteristic profiles derived in the study by \citet{2002A&A...390.1089P} are shown at matched resolution in grey.}
  \label{fig:fig2}
\end{figure}

\begin{figure}[htb]
  \centering
  \includegraphics[angle=0, width=\linewidth]{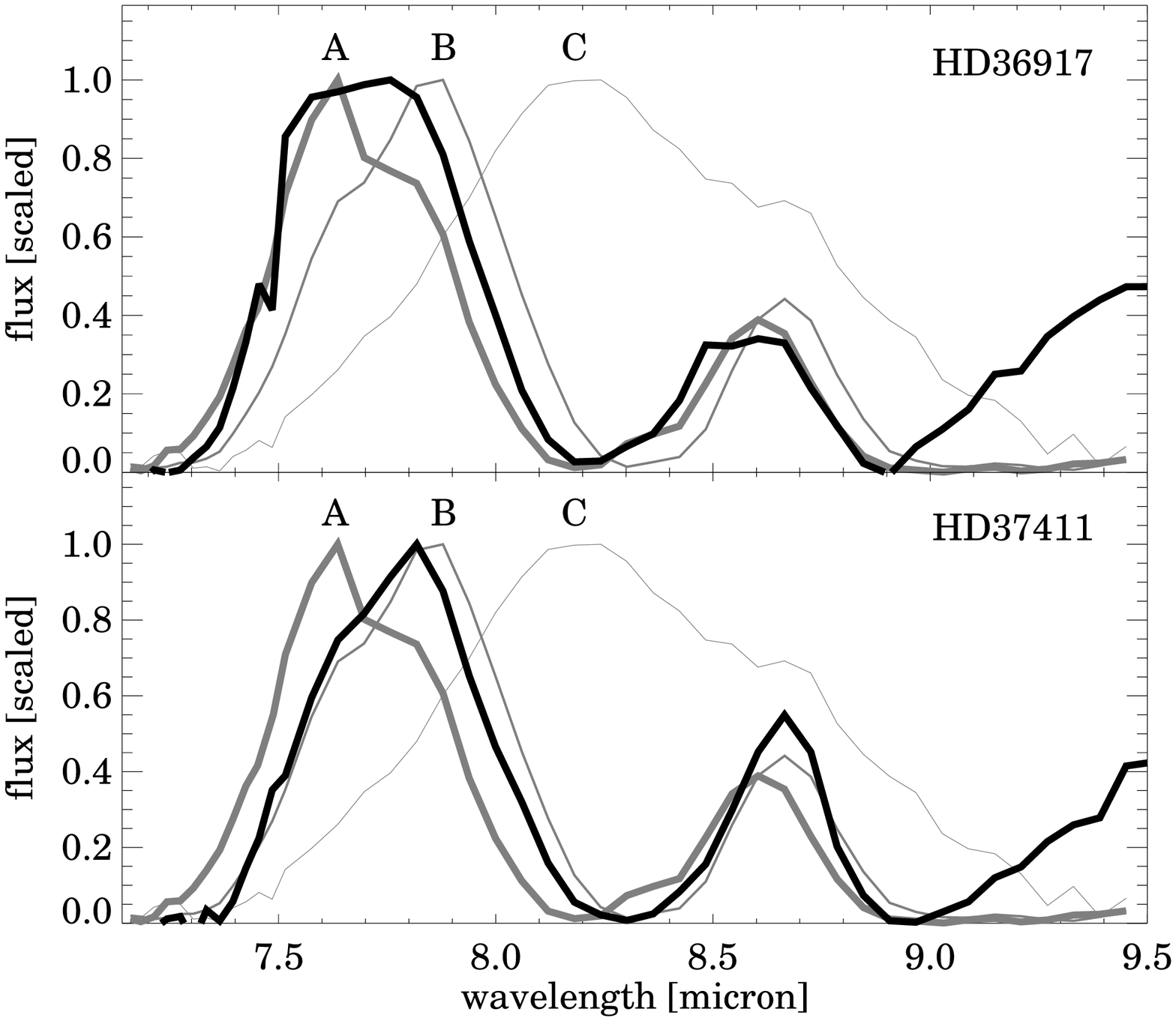}
  \caption{The derived ``7.7'' $\mu$m profiles (black) after subtraction of the underlying continuum. For comparison, the three characteristic profiles derived in the study by \citet{2002A&A...390.1089P} are shown at matched resolution in grey.}
   \label{fig:fig3}
\end{figure}

\begin{figure*}[htb]
  \centering
  \includegraphics[angle=90, width=\linewidth]{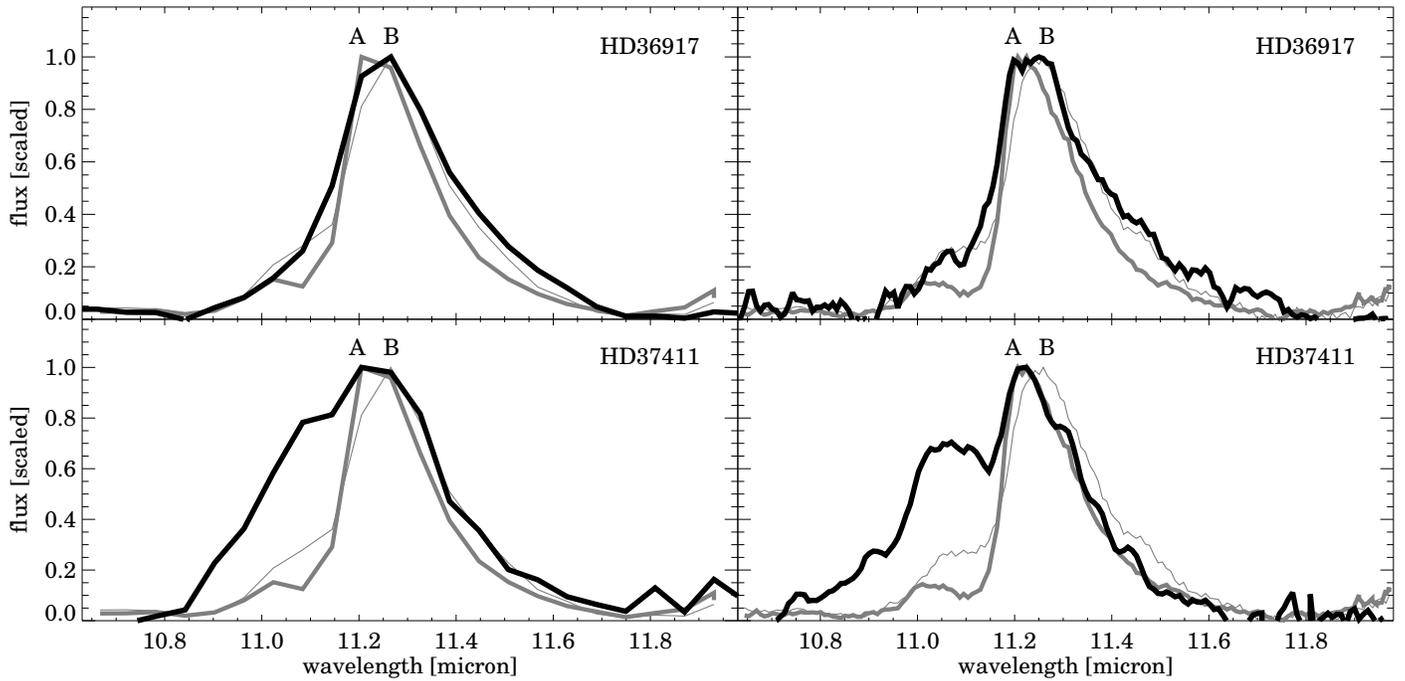}
  \caption{The derived 11.2 $\mu$m profiles (black) after subtraction of the underlying continuum and correction for the broad 10 $\mu$m feature in HD37411. For comparison, the two characteristic profiles derived in the study by \citet{2004ApJ...611..928V} are shown at matched resolution in grey. Left: The derived low-resolution (SL) profiles. Right: The derived high-resolution (SH) profiles.}
  \label{fig:fig4}
\end{figure*}

The apparent variations in peak position, width and wing shapes
between the derived profiles is much larger than could be caused by
the uncertainty in the adopted continuum. Studies of the detailed
profiles of the PAH bands in the ISO SWS spectra of a large sample of
stellar sources, planetary nebulae (PNe), reflection nebulae,
\ion{H}{ii} regions and galaxies have revealed that the bands in the 6
- 9 $\mu$m range show strong variation in peak position and profile
\citep{2002A&A...390.1089P}. Those studies showed that the variations
in the PAH bands correlate with object type. All ISM-like sources
belong to group A characterised by a ``6.2 $\mu$m'' band peaking at
6.2 $\mu$m and a 7.7 $\mu$m band peaking at 7.6 $\mu$m, and while the
isolated Herbig Ae/Be stars, along with a few post-AGB stars and most
PNe, belong to group B, which are characterised by a ``6.2'' $\mu$m
band peaking around 6.28 $\mu$m and a 7.7 $\mu$m band peaking at 7.9
$\mu$m. Finally, two post-AGB stars make up group C, characterised by
a ``6.2'' $\mu$m band peaking at 6.3 $\mu$m and with no apparent 7.7
$\mu$m band, but a broad ``8.22'' $\mu$m feature (Fig. \ref{fig:fig2}
- \ref{fig:fig4}; Table \ref{tab:tab6}). In
\citet{2004ApJ...611..928V}, the variations in the 3.3 and 11.2 $\mu$m
bands are studied. However, the observed variation are shown to be
more modest and the object correlation not as tight than those
observed for the 6 - 9 $\mu$m region.

The observed diversity in peak position, profile, and relative
strength of the PAH bands are ascribed to global changes in the
physical and chemical characteristics of the emitting PAH family. More
specifically, variations in the peak positions of the 6.2 $\mu$m band
are thought to indicate incorporation of nitrogen in the aromatic ring
structure \citep{2002A&A...390.1089P,2002ApJ...564..782B}. The 7.6/7.8
$\mu$m bands are also likely to be related to chemical modifications,
based on the strong correlation with the 6.2 $\mu$m variations. In
contrast, variations in the ratio of the C-H modes to the C-C modes
(e.g., 11.2/6.2 $\mu$m) are attributed to variations in the charge
state of the emitting PAH molecules
\citep{1999A&A...352..659A,2001A&A...370.1030H,2001ApJ...556..501B}.
Emission between 10.9 and 11.1 $\mu$m is attributed to the
out-of-plane bending vibrations of solo-CH groups on the periphery of
moderately sized ($\sim$100 carbon atoms) PAHs
\citep{1999ApJ...516L..41H}. In Fig. \ref{fig:fig2} - \ref{fig:fig4},
the derived profiles from the two Spitzer sources are compared to the
characteristic profiles from \citet{2002A&A...390.1089P} and
\citet{2004ApJ...611..928V}.

\begin{table}[htb]
  \centering
  \caption{The spectral characteristics for the classes from \citet{2002A&A...390.1089P} and \citet{2004ApJ...611..928V}. $\lambda_{x}$ indicates the peak positions, in $\mu$m.}
    \label{tab:tab6}
    \begin{minipage}{\linewidth}
      \centering
      \begin{tabular}{@{}p{0.7cm}p{2.6cm}ccc@{}}
        \hline
        \hline
        \multicolumn{5}{c}{\bf The 5 - 9 $\mu$m region} \\
        Class          & Object                 & $\lambda_{6.2}$ &  $\lambda_{7.7}$\footnotemark[1] &  $\lambda_{8.6}$ \\
        \hline                                     
        $\mathcal{A}$  & ISM-like               & $\sim$6.22     & 7.6/equal                        & $\sim$8.6       \\
        $\mathcal{B}$  & non-ISM-like           & 6.24 - 6.28    & ``7.8''                          & $>$8.62         \\
        $\mathcal{C}$  & non-ISM-like           & $\sim$6.3      & 8.22                             & none            \\ 
        \hline
        \hline
        \multicolumn{5}{c}{\bf The 11.2 $\mu$m region} \\
        Class         & Object                  & $\lambda_{11.2}$           & \multicolumn{2}{c}{FWHM$_{11.2}$} \\  
        \hline
        A$_{11.2}$     & ISM-like                & $\sim$ 11.20 - 11.24      & \multicolumn{2}{c}{$\sim0.17$}    \\
        A(B)$_{11.2}$  & uncorrelated            & $\sim$ 11.20 - 11.24      & \multicolumn{2}{c}{$\sim0.21$}     \\
        B$_{11.2}$     & uncorrelated            & $\sim$ 11.25              & \multicolumn{2}{c}{$\sim0.20$}     \\
        \hline
      \end{tabular}
      \begin{flushleft}
        1 : The ``7.7'' $\mu$m feature is classified by its dominant component, the 7.6 and/or the 7.8 $\mu$m component.
      \end{flushleft}  
    \end{minipage}
\end{table}

In Fig. \ref{fig:fig2}, the 6.2 $\mu$m profiles are compared. The 6.2
$\mu$m profile of HD36917 resemblances that of class A best and for
HD37411 that of class B. The profiles in both classes show a variation
in the strength of the red tail, relative to the peak. In that respect
the red tails of both sources fall off more rapidly than the
characteristic profile that defines their class, but it is known to
vary \citep{2002A&A...390.1089P}.

The variations in the ``7.7'' $\mu$m feature (Fig. \ref{fig:fig3}) are
particular interesting. Remarkably, HD36917 is best seen as a
\emph{blend} of a class A and class B profile. The ``7.7'' $\mu$m band
in the spectrum of HD37411 is a clear class B profile.

The differences between the 11.2 $\mu$m profiles are subtle and
difficult to discern, despite the high-resolution profiles (Fig.
\ref{fig:fig4}). Again, HD36917 shows evidence for a blend of a class
A and class B profile. The classification for HD37411 is hampered by
the dominating continuum component between 9 - 13 $\mu$m (inset Fig.
\ref{fig:fig1}). As described in Sect. \ref{subsec:sec3subsec1}, this
additional component has been approximated by a single spline and has
been subtracted. The result is presented in Fig. \ref{fig:fig4}

Both spectra show evidence for a 11.1 $\mu$m feature. Many sources
show this weak `satellite' accompanying the 11.2 $\mu$m feature
\citep{2001A&A...370.1030H}. However, the 11.1 $\mu$m feature in
HD37411 is very strong relative to the 11.2 $\mu$m band. This spectral
region is often confused by emission bands due to forsterite. In
astronomical spectra, these often coincide with the 11.2 $\mu$m band.
However, we note that laboratory spectra show evidence for forsterite
near 11.0 $\mu$m \citep{2006ApJ...648L.147T}. The broad 10 $\mu$m
band, evident in Fig. \ref{fig:fig1}, indicates the presence of warm
silicate in the source and supports this suggestion. In any case,
focussing on the PAHs and guided by the high-resolution data, the
profile is classified A. Table \ref{tab:tab7} summarises the assigned
classes.

\begin{table}[htbp]
  \centering
  \caption{Assigned classes, see text for details.}
  \label{tab:tab7}
  \begin{minipage}{\linewidth}
    \centering
    \begin{tabular}{llll}
      \hline
      \hline
                 & 6.2 & 7.7   & 11.2    \\
      \hline
      HD36917    & A   & A + B & B (+ A?)\\
      HD37411    & B   & B     & A       \\
      \hline
      TY CrA     & A   & A     & A       \\ 
      HD97048    & A   & A + B & A       \\ 
      HD100546   & B   & B     & -\footnotemark[1]\\ 
      \hline
    \end{tabular}
    \begin{flushleft}
      1 : Due to strong silicate contamination, classification of the 11.2 $\mu$m profile has been omitted.
    \end{flushleft}
  \end{minipage}
\end{table}

\section{Discussion}
\label{sec:sec4}

\cite{2007ApJ...664.1144S} have identified systematic variations in
the peak position of the 7.7 and 11.2 $\mu$m band with the spectral
type of the exciting star. We also recognise such variations in the
spectra of HD39917 and HD37411. We note that both stars have similar
spectral type ($\sim$B9). Going back to ISO SWS
\citep{1996A&A...315L..49D} data \citep{2002PhdT........D} we have
retrieved additional spectra of three well-studied and
well-characterised Herbig Ae/Be stars, which provide additional
insight in the possible factors driving these spectral variations.

The first star is TY CrA, a double-lined eclipsing binary consisting
of a Herbig Ae/Be object \citep{1994A&AS..104..315T}, close to the
zero age main sequence, and a late-type pre-main-sequence object. The
object is located near one of the densest parts of the R Corona
Australis star-forming region, embedded in the bright reflection
nebula NGC6726/6727 \citep{1993A&A...278..569H}. The infrared emission
seen in the ISO spectrum does not originate from TY CrA itself, but
from the nearby `TY CrA bar' \citep{2000A&A...361..258S,
  2007A&A...476..279G}. The second star is HD97048, it is located in
the Chamaeleon I cloud of the Chamaeleon T association, which
illuminates the bright reflection nebula Ced 111. Spatial studies on
the PAH emission done with the VLT's spectrometer and imager for the
mid-infrared \citep[VISIR;][]{2004Msngr.117...12L} and ISO have
revealed a circumstellar ($\sim0.5^{\prime\prime}$) and a more
extended component ($>5.4^{\prime\prime}$;
\citealt{1994A&A...292..593P,
  2000A&A...361..258S,2004A&A...418..177V,
  2006A&A...449.1067H}). Finally, HD100546 is a typical isolated
Herbig Ae/Be star, which has no companions within 1500 AU and is not
associated with the nearby dark cloud DC 292.6-7.9
\citep{2001AJ....122.3396G}. The mid-IR emission of this source
clearly originates from the circumstellar protoplanetary disk, which
has a gap of about $\sim$10 AU with a `puffed-up' inner rim
\citep{2006A&A...449..267A}.

\subsection{The profiles}
\label{subsec:sec4subsec1}

\begin{figure*}
  \centering
  \includegraphics[angle=90, width=\linewidth]{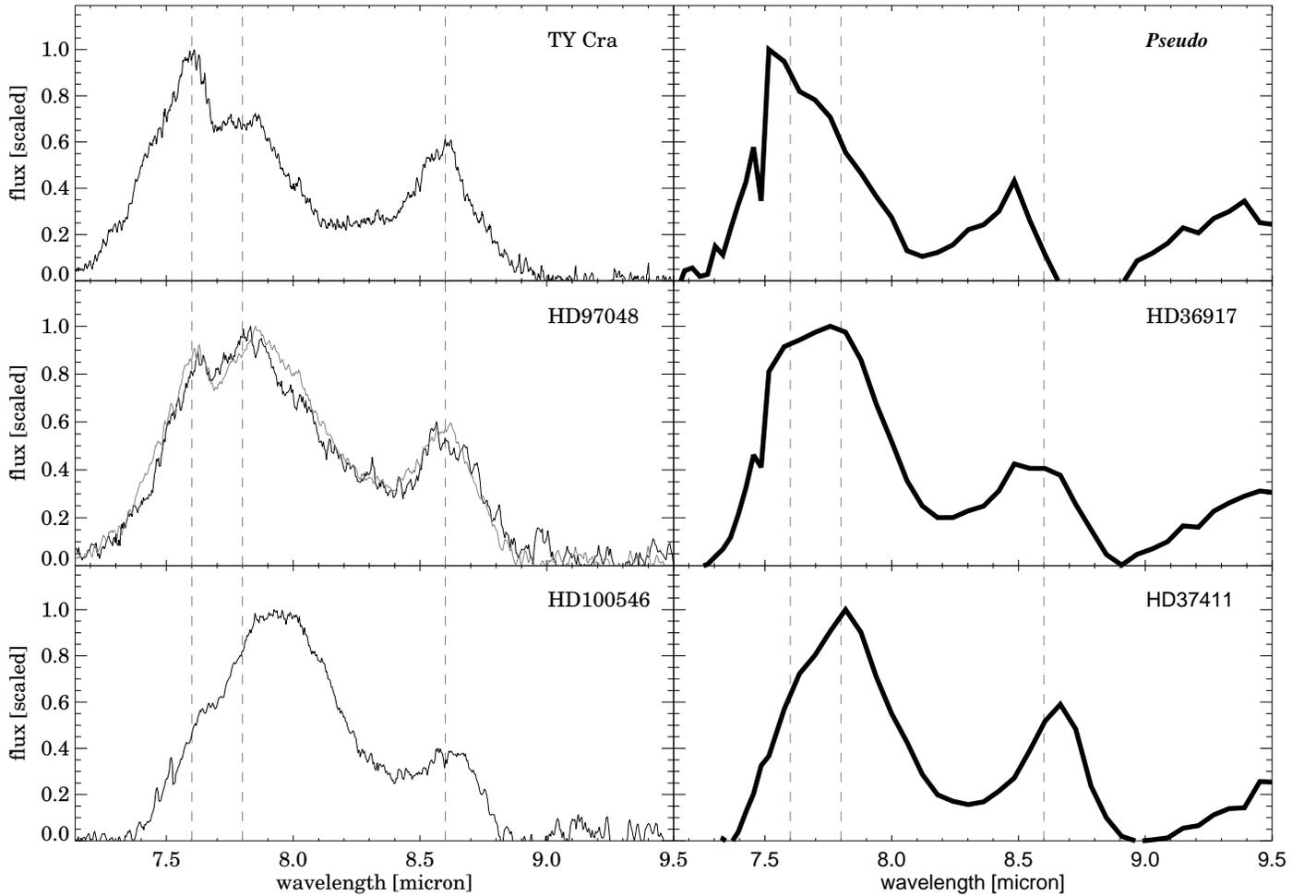}
  \caption{Left: The normalised profiles of the ``7.7'' $\mu$m feature of HD97048, TY CrA, and HD100546, as obtained by the SWS instrument on board ISO. The dashed lines indicate the peak positions of the major PAH band components. In grey the combined profile for HD97048, constructed from TY CrA and HD100546, is presented. Note the quality of the match. Right: The normalised profiles of the ``7.7'' $\mu$m feature of HD36917 and HD37411, as obtained by the IRS instrument on board Spitzer. The profile labled {\it pseudo} is obtained by subtracting the profile of HD37411 from that of HD36917. The dashed lines indicate the peak positions of the major PAH band components.}
  \label{fig:fig5}
\end{figure*}

Figure \ref{fig:fig5} compares the ``7.7'' $\mu$m feature
profiles\footnote{For details concerning the ISO SWS spectra, such as
  data reduction and analyses we refer to \cite{2002PhdT........D}.}
of all five sources. The peak position of the 7.7 $\mu$m feature
varies considerably in this sample from 7.6 $\mu$m all the way to 8.0
$\mu$m. Most striking is that the profile for HD97048 can be obtained
by combining the profiles from TY CrA and HD100546
\citep{2002PhdT........D}. The two \emph{Spitzer} sources show similar
behaviour. Indeed, the 7.7 $\mu$m feature of HD36917 may be obtained
by combining the class B profile of HD37411 with a class A-like
profile (labeled {\it pseudo}) in Fig. \ref{fig:fig5}.

Table \ref{tab:tab7} shows that the systematic differences between the
profiles of the three \emph{ISO} sources are also analog to those
between the profiles of the two \emph{Spitzer} sources. The spectral
type of the five sources are similar ($\sim$B9), therefore, the
variations in the derived profiles between these stars cannot be
caused by differences in the exciting spectrum.

\begin{figure}
  \centering
  \includegraphics[angle=90, width=\linewidth]{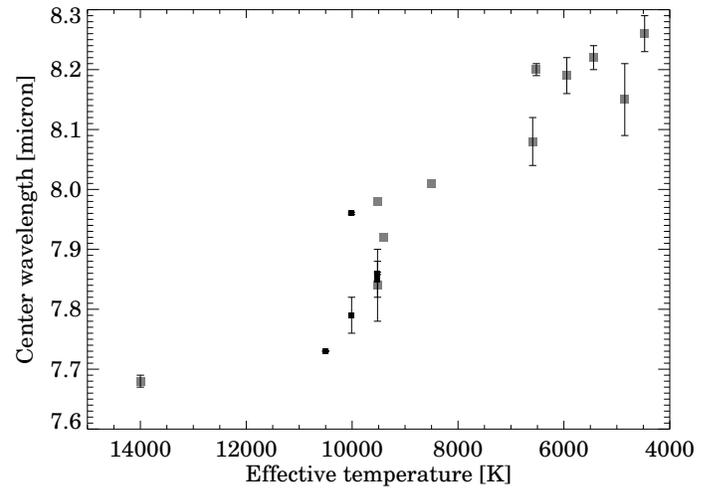}
  \caption{Correlation between the central wavelength of the 7 - 8 $\mu$m band and the effective temperature of the exciting star. The 5 Herbig stars used in this study are shown as black squares. Adopted from \cite{2007ApJ...664.1144S}}
  \label{fig:fig6}
\end{figure}

\cite{2007ApJ...664.1144S} find a correlation between the central
wavelength of the 7 - 8 $\mu$m feature and the effective temperature
of the exciting star. We have determined the peak position of the 7 -
8 $\mu$m band for the 2 \emph{Spitzer} and 3 \emph{ISO} sources in our
study in the same manner as \cite{2007ApJ...664.1144S} and plotted our
results with the data from
\cite{2005ApJ...632..956S,2007ApJ...664.1144S} in Fig. \ref{fig:fig6}.
Although there is no clear break in the correlation, a jump is
discernible around 10,000 Kelvin. Following
\cite{2007ApJ...664.1144S}, the diagram is interpreted to reveal a
``7.7'' $\mu$m band peaking at 7.6 $\mu$m (class A) for
T$_{\rm eff}\gtrsim$ 10,000 K, at 7.9 $\mu$m (class B) for 10,000
$\gtrsim$ T$_{\rm eff}\gtrsim$ 7000 K and at 8.2 $\mu$m (class C) for
T$_{\rm eff}\lesssim$ 7000 K. We note that this subdivision parallels
a subdivision in source classification as with class A containing
sources dominated by ISM material, class B comprising Herbig stars,
and class C consisting of post-AGB objects. We point out that in this
sample objects, like the PNe NGC 7027, with T$_{\rm eff}\simeq$
200,000 K and belonging to \emph{class B}, are not shown.

\subsection{Extended PAH band emission}
\label{subsec:sec4subsec2}

\citet{2002PhdT........D} and \citet{2003asdu.confE..41P} note that
the observed variations in the 7 - 9 $\mu$m region might well be due
to differences in morphology of the sources, combining ISM material
and/or disks. Most resounding in this respect is the source HD97048,
which morphologically consists of a spatial combination of
circumstellar disk material and interstellar material
\citep{2004Msngr.117...12L} and spectroscopically combines
characteristics of classes A and B.

Linking back to the two \emph{Spitzer} sources, this connects the
detected PAH emission from HD37411 to the circumstellar disk and,
particularly fascinating, the PAH emission from HD36917 to, in part,
the surrounding cloud and/or envelope (pseudo source) and, in part,
the circumstellar disk. Intrigued by the notion of two different,
spatially, separated PAH families in HD36917, we are motivated to
investigate the \emph{spatial} variation of the PAH emission coming
from HD36917.

\begin{figure}[htb]
  \centering
  \includegraphics[angle=0, width=\linewidth]{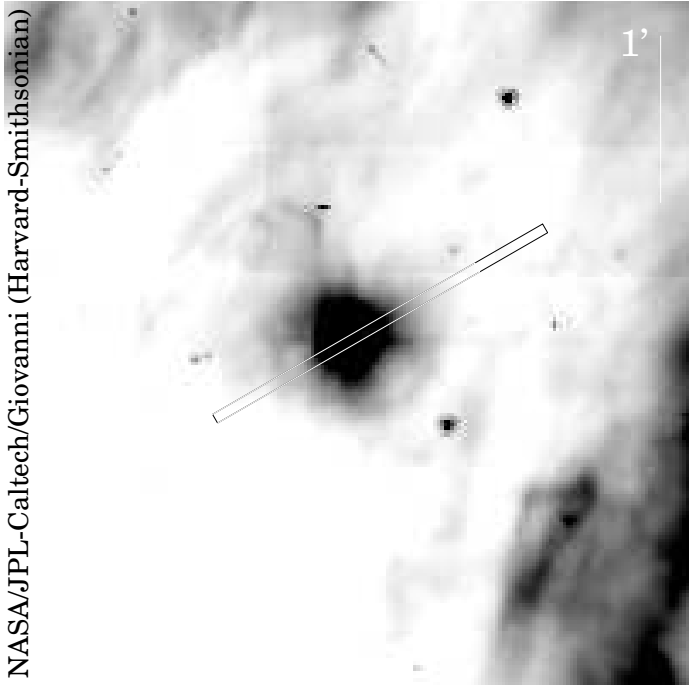}
  \caption{8 $\mu$m IRAC mosaic from part of the OB1c association, centred on HD36917 and the colour scaling is chosen to emphasise the emission coming from the envelope. Also shown is the IRS SL slit on the source.}
   \label{fig:fig7}
\end{figure}

The available 8 $\mu$m IRAC \citep{2004ApJS..154...10F} mosaic from
part of the OB1c association is presented in Fig. \ref{fig:fig7}. The
image is centred on the position of HD36917 and the colour scaling is
chosen to emphasise the emission from the envelope. The strong and
rich morphology of the infrared background is clearly visible. The
IRAC data suggest diffuse extended nebulosity associated with the
source on scales up to at least $\sim1^{\prime}\ (3\cdot10^{4}\
\mathrm{AU})$.

\begin{figure}[htb]
  \centering
  \includegraphics[angle=0, width=\linewidth]{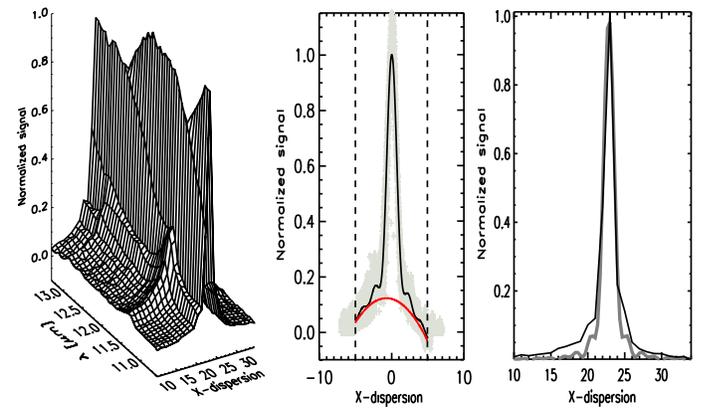}
  \caption{Left: The normalised source profile of a section from the SL1 long-slit 2-D spectral image as observed toward HD36917. As a result of the angle of the order across the image, the profile centre moves in cross-dispersion. This results in a changing distribution of the profile signal across pixels, hence the apparent lower than unity profile at certain wavelengths. The presence of a spatially extended component near the 11.2 and 12.7 $\mu$m PAH bands in HD36917 is clearly visible. Middle : The simultaneous fit (black) of a source profile and a 2$^{\rm 2n}$ order polynomial to the combined dither data (grey). Right: The average cross-dispersion profile from 11.2 to 11.5 $\mu$m is compared with the calibration profile derived from the standard stars. For the extraction the calibration cross-dispersion profile is fit in combination with a $2^{\rm nd}$ order polynomial to decompose the spatial components.}
  \label{fig:fig8}
\end{figure}

\begin{figure}[htb]
  \centering
  \includegraphics[width=\linewidth]{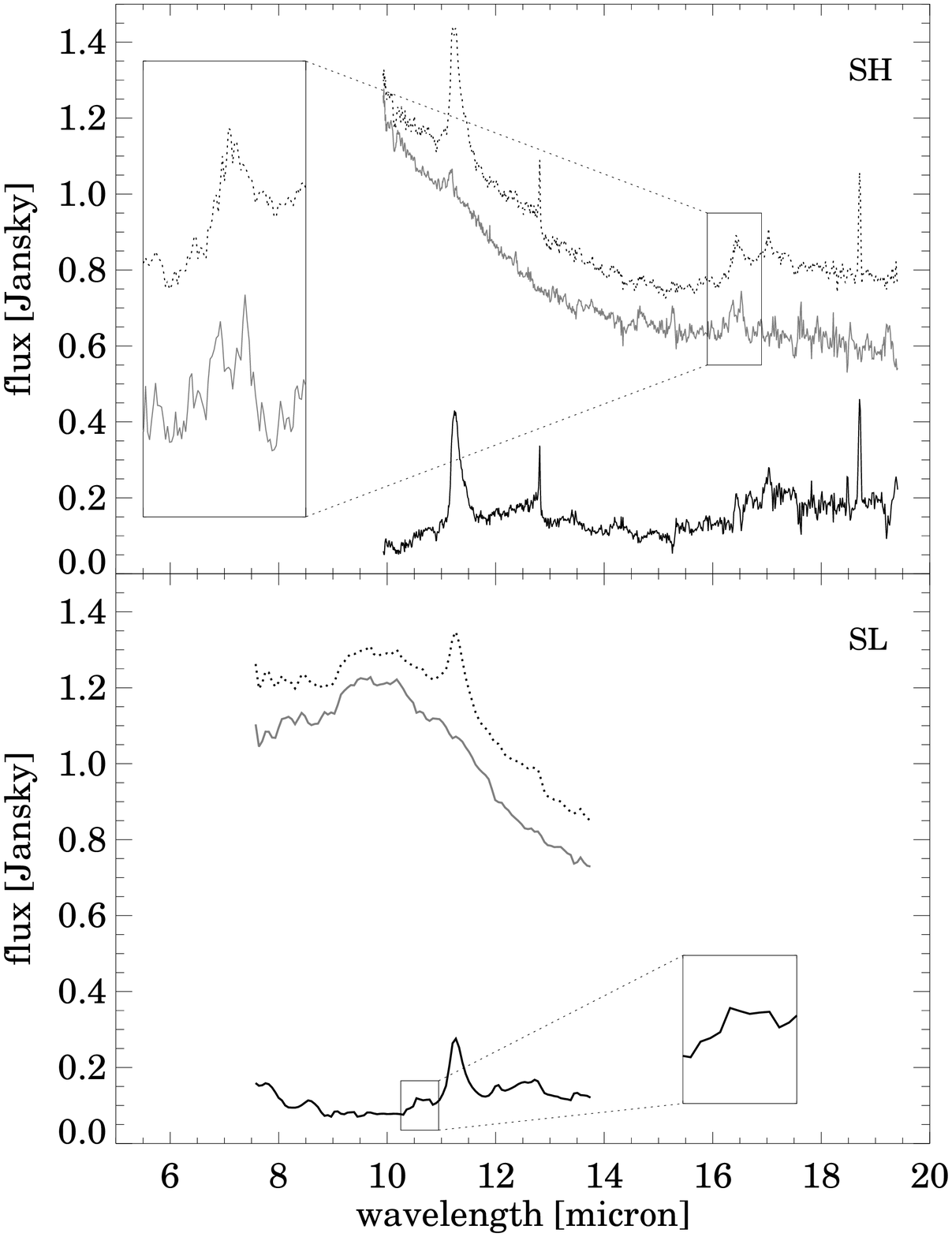}
  \caption{The local background corrected SL (bottom) and SH (top) spectra for HD36917, as obtained with the decomposition. The corrected spectra are grey, the uncorrected spectra are dotted grey and the background estimates are black. The boxes show a few interesting spectral regions.}
  \label{fig:fig9}
\end{figure}

Fig. \ref{fig:fig8} illustrates the use of the source profile
extraction to decompose both spatial components (see Sect.
\ref{sec:sec2}). In the left panel, part of the first order in the SL
module of the spectral image of HD36917 is shown, demonstrating the
extended nature of the 11.2 and 12.7 $\mu$m PAH bands in HD36917. The
panel in the middle displays the result (black) of simultaneous
fitting a source profile and a 2$^{\rm nd}$ order polynomial (red) to
the combined dither data (grey). A $2^{\rm nd}$ order polynomial for
the extended emission produces the optimal extraction, providing a
reasonable approximation to the observed emission. A higher order
polynomial results in increased noise, while a lower order polynomial
provides an inferior fit and underestimates the extended emission
component. The half-width of the extended emission is about 6 pixels
($\simeq11^{\prime\prime}\ ;\ 6\cdot10^{3}\ \mathrm{AU}$). The panel
to the right in Fig. \ref{fig:fig8} shows the cross-dispersion profile
from 11.2 to 11.5 $\mu$m (black) compared with the calibration profile
derived from standard stars (grey). In Fig. \ref{fig:fig9}, the SL and
SH spectra, corrected for local background emission, are presented.

Remarkably, in both the SL and SH spectra the PAH signatures have
largely disappeared. As expected, this procedure removes the forbidden
atomic lines in the SH spectrum. In addition, the 11.2, and 12.7
$\mu$m band and the plateau underneath 16.4 $\mu$m have vanished.
Obviously, the red end of the ``7.7'' $\mu$m feature, the 11.2 and
12.7 $\mu$m PAH band emission originates from the surrounding
cloud/envelope. It is clear that in this source there is \emph{no}
evidence for PAH emission associated with the direct (unresolved)
circumstellar disk ($\sim0.2^{\prime\prime}\ ;\ R\sim50\
\mathrm{AU}$).

Note that, in contrast to the PAH features, the 16.4 $\mu$m band seems
to be associated with the direct circumstellar environment of the
star. Although the band is located in a region where two orders
overlap (12 and 13) and, therefore, the extended emission estimate is
somewhat more uncertain, forsterite does show some weak features near
this position \citep[16.3 $\mu$m]{2003A&A...399.1101K}. This actually
suggests the presence of processed silicate material in the disk.
Furthermore, spectral structure between 10.5 - 10.9 $\mu$m appears in
the background corrected spectra, most distinctly seen in the SL
spectrum. This could be the weak PAH band first recognised by
\cite{1989ApJ...341..270W} and later, e.g., \cite{1996A&A...315L.369B}.

The absence of PAH features in the spectra of the disk is not too
surprising. \cite{2001A&A...365..476M} classified 46 spectra of Herbig
Ae/Be stars, based on the shape of the SED in the mid-IR, in
distinctive groups. Their group I sources all showed emission due to
PAHs, but their group II sources show no or only weak PAH emission.
They link this difference to the geometry of the circumstellar disk;
flaring vs. self-shadowed. The SED of HD36917 makes it a Group II
source \citep{2004A&A...427..179H}, consistent with this
interpretation.

The PAH emission in HD36917 originates from a region $6\cdot10^{3}$ AU
in size and, therefore, is likely associated with an extended envelope
rather than a disk. Thus, although the spectra show evidence for a
`processed' PAH component peaking toward the source - thus associated
with the source - the processing itself seems to occur on a
scale-size, which is much larger than a star/disk system. Therefore,
it has the appearance that the spectral variations can be attributed
to the spatial structure of the region, although division into a clear
disk and cloud component is not evident from our data.

\subsection{Origin of the variations}
\label{subsec:sec4subsec3}

The observed trend between the central wavelength of the 7.7 - 7.9
$\mu$m feature and the spectral type of the irradiation star, as shown
in Fig. \ref{fig:fig6}, shows a jump near 10,000 Kelvin. The spectral
variations in the ``7.7'' $\mu$m feature reflect chemical
modifications and \cite{2007ApJ...664.1144S} suggested that these
modifications are likely driven by UV processing. But hot class B
sources, such as NGC 7027, seem to indicate that other source
characteristics, besides effective temperature, may be of importance
too. Such characteristics could include local density or history of
the carriers. Now, the jump near 10,000 Kelvin seems to indicate that
for Herbig Ae stars these other factors are of consequence. In this
study the spectral variations are linked to PAHs in spatial structure,
promoting differences in UV processing. Note that the extended nature
of HD36917 lends itself particularly well to a spatial-spectral study
and may well settle the factors involved in the chemical processing of
PAHs.

\subsection{Implications}
\label{subsec:sec4subsec4}

The chemical modifications driving the observed variations in Herbig
Ae stars may well be due to a transition from stable aromatic
structures in the cloud (class A), to more labile aliphatic-like
structures in the circumstellar environment (class B)
\citep{2007ApJ...664.1144S}. However, whereas post-AGB objects (class
C) and ISM sources (class A) lend themselves to an evolutionary
interpretation - from labile aliphatic hydrocarbons in the benign
conditions of post-AGB objects to stable aromatic hydrocarbons in the
harsh environment of the ISM - it would be unlikely for the aromatic
structures to regenerate their aliphatic side-groups when going from
the ISM to the protoplanetary disks. Rather, we interpret the observed
spectral variations as reflecting the presence of an active chemical
balance - in \emph{all} sources - between hydrogenation, carbon
reactions building (aliphatic) hydrocarbons, and UV photons breaking
them down.

\section{Summary and conclusions}
\label{sec:sec5}
From a sample of 15 Herbig Ae/Be stars we analysed the two PAH
dominated \emph{Spitzer} IRS spectra. The derived profiles have been
classified according the scheme of \cite{2002A&A...390.1089P}.
Comparison of these profiles with profiles obtained using \emph{ISO}'s
SWS from three (spatially) well studied objects indicate the presence
of two dissimilar, spatially separated, PAH families for HD36917.

The analysis presented here shows again, based on the variation in
band profiles, that PAHs in subsequent, evolutionary linked stages of
star formation, are different from those in the general ISM
\citep{2001A&A...370.1030H}. While previously the required chemical
processing of the PAHs was placed in the protoplanetary disk phase
\citep{2002PhdT........D, 2003asdu.confE..41P}, the evidence presented
here, shows that such chemical changes may already occur in the
(collapsing?) natal cloud. The supporting evidence for this is that
the detected PAH emission cannot be associated with the (unresolved)
disk and must thus be associated with the circumstellar (natal) cloud.
Furthermore, the presence of the two distinct PAH families suggests
active chemistry. Further studies on the spatial and spectral
variations in the PAH spectra of these sources may be helpful to
disentangle the factors that stimulate the processing of PAH
molecules.

Future observation, using (high-resolution) spectral mapping
techniques, should provide a better handle on the spatial variations
of the PAH emission/profiles coming from and surrounding HD36917 and
possibly link them to the changing radiation field as one moves out
from the central star. Additional laboratory studies are required to
explain the transition from class A PAH profiles in the circumstellar
(natal) cloud to class B profiles in disks.

\begin{acknowledgement}
  We gratefully thank Els Peeters for providing the ISO SWS data and
  acknowledge Mario van den Ancker for providing a literature overview
  on the Herbig Ae/Be stars investigated.
\end{acknowledgement}

\bibliographystyle{aa}
\bibliography{aamnem99,bibliography}

\end{document}